\documentclass{elsart}
\usepackage{graphics}
\usepackage{graphicx}
\usepackage{epsfig}
\usepackage{amssymb}

\begin{document}

\begin{frontmatter}

\title{HERA-B Framework for Online Calibration and Alignment}

\author[desy-HH,desy-ifh]{J.M. Hern\'andez\thanksref{ciemat},}
\author[desy-HH]{D. Ressing,}
\author[desy-HH]{V. Rybnikov,}
\author[desy-HH]{F. S\'anchez\thanksref{ifae},}
\author[lip]{A. Amorim,}
\author[desy-HH]{M. Medinnis,}
\author[desy-HH]{P. Kreuzer\thanksref{athens},}
\author[desy-ifh]{U. Schwanke\thanksref{hu}}
\address[desy-HH]{DESY, D-22603 Hamburg, Germany}
\address[desy-ifh]{DESY, D-15738 Zeuthen, Germany}
\address[lip]{FCUL and LIP, P-1749-016 Lisboa, Portugal}
\thanks[ciemat]{Now at CIEMAT, E-28040 Madrid, Spain}
\thanks[ifae]{Now at Universitat Aut\`onoma Barcelona/IFAE, E-08193 Bellaterra, Spain}
\thanks[athens]{Now at Athens University, Athens, Greece}
\thanks[hu]{Now at Humboldt University, Berlin, Germany}

\begin{abstract}
  This paper describes the architecture and implementation of the
  HERA-B framework for online calibration and alignment.  At HERA-B the
  performance of all trigger levels, including the online
  reconstruction, strongly depends on using the appropriate
  calibration and alignment constants, which might change during data
  taking. A system to monitor, recompute and distribute those
  constants to online processes has been integrated in the data
  acquisition and trigger systems.

\end{abstract}

\begin{keyword}
Conditions database \sep calibration \sep alignment \sep online reconstruction \sep PC farms
\PACS \\
07.05.-t  Computers in experimental physics \\
07.05.Hd  Data acquisition: hardware and software\\
07.05.Bx  Computer systems: hardware, operating systems, computer languages and utilities \\

\end{keyword}
\end{frontmatter}

\section{Introduction}
\label{introduction}

It is essential in High Energy Physics experiments, that accurate and
consistent detector parameter sets are used at all trigger levels and
also in the event reconstruction.  Similarly, simulation programs must
use detector parameters which are consistent with those used in the
trigger and reconstruction to properly simulate the detector and
trigger conditions. The detector parameters (such as calibrations,
alignments, detector channel maps, resolutions, etc), globally known
as detector conditions\footnote{The detector conditions will be
  hereinafter also referred as CnA ({\underline C}alibration
  a{\underline n}d {\underline A}lignment) constants}, are normally
calculated offline in a sporadic manner from monitoring information
and event data, and updated in the trigger and reconstruction codes.
The bookkeeping of the detector conditions becomes then an important
issue.

The HERA-B experiment \cite{herab} was designed for the measurement of
CP violation in the neutral B-meson system. The data acquisition (DAQ)
and trigger systems were designed to cope with more than half a
million detector channels, a 40 MHz interaction rate and an extremely
low signal to background ratio of $10^{-10}$. A networked
high-bandwidth data acquisition system \cite{daq} and a highly
selective multi-level trigger \cite{trigger,hlt}, with a suppression
factor of $10^{6}$, were built.  Unlike most HEP experiments, HERA-B
performs full event reconstruction online.


A novel approach for handling the detector conditions has been
followed at HERA-B where a system to monitor, recompute and distribute
CnA constants to online clients is integrated into the DAQ and trigger
systems. Online updates of the CnA constants help to stabilize trigger
performance and online reconstruction as detector conditions vary during
data taking.
The CnA system is also employed offline during event data reprocessing
and Monte Carlo reconstruction. It allows to incorporate offline
updates of the CnA constants during the data reprocessing and also
ensures that the reconstruction of Monte Carlo simulated events is
performed using the same CnA constants employed in the reconstruction
of the real data being simulated.  This approach is of potential
interest to future HEP experiments who are planning sophisticated
trigger systems and online event reconstruction.

The architecture, implementation and performance of the CnA system are
described in this paper.  The motivation for the CnA system and its
requirements are summarized in the next section. The system
architecture is described in section~\ref{s:design} and its
implementation and performance is discussed in section~\ref{s:desc}.
Section~\ref{s:mc} describes the offline usage of the CnA system for
data reprocessing and Monte Carlo reconstruction.

\section{Motivation and requirements}

The design and requirements of the online CnA system are driven by the
design of the HERA-B detector and the architecture of the DAQ and
trigger systems.  We therefore begin this section with a description
of the relevant aspects of the HERA-B detector, DAQ and trigger
systems.

\begin{figure}[htp]
\centering
\includegraphics[width=\textwidth]{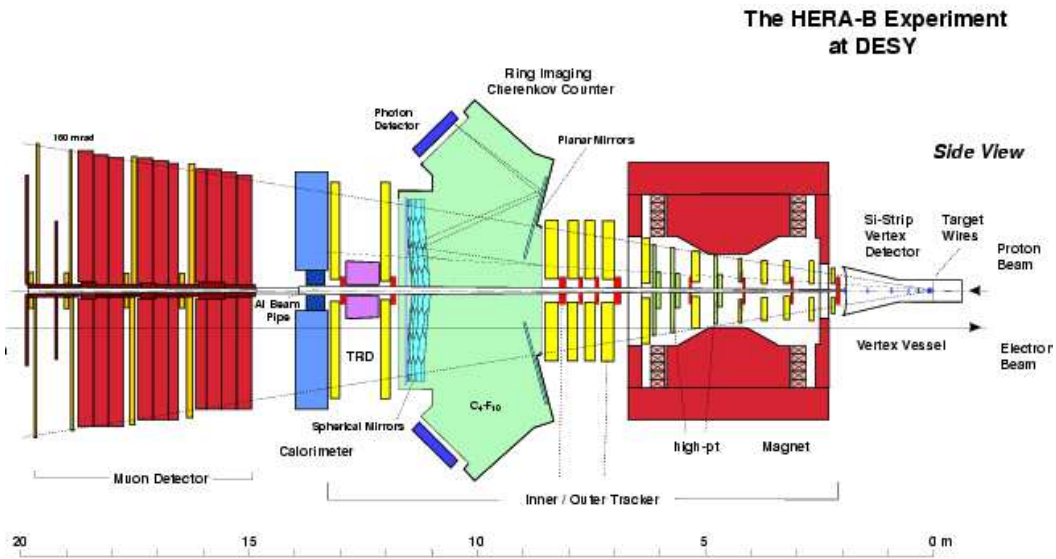}
\caption{Side view of the HERA-B detector}
\label{fig:detector}
\end{figure}

The HERA-B detector is depicted in figure~\ref{fig:detector}.  The
target wires and the silicon vertex detector (SVD) stations are
movable. The target positions change to
stabilize the average interaction rate.
The SVD stations are also moved towards the beam at the beginning of
every fill and retracted at the end to avoid damage during injection.
These two subsystems are particularly subject to alignment changes.
Although the stepping motors provide sufficient precision (of order
1~micron), alignment corrections are needed relatively often for
optimal performance due to thermal and other effects. Similarly, the
realignment of the tracking chambers is needed whenever they are moved
away from the beam line during accesses to the detector for repairs.
Such accesses occur typically at intervals of order one week.  The
calibration of all detector subsystems is often updated as well.
Examples are pedestal following and energy calibration of the
electromagnetic calorimeter, the time calibration of the TDC boards of
the tracker system, the drift velocity calibration of the tracker
system and the maximal value of the Cherenkov angle in the RICH
detector which varies with atmospheric pressure, temperature and gas
composition.  Moreover, the channel status (hot/dead/noisy) maps for
all subsystems must be monitored and updated periodically.

\begin{figure}[htp]
\centering
\includegraphics[width=0.8\textwidth]{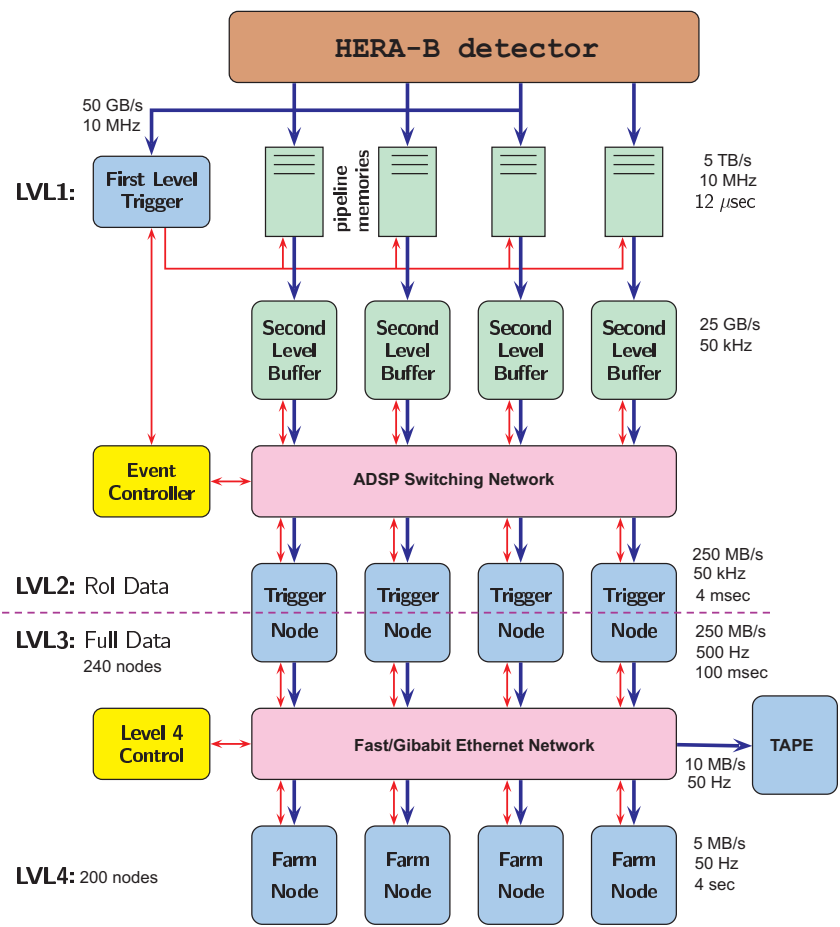}
\caption{Scheme of the data acquisition and trigger systems. The data throughput,
  trigger rates and latencies for each of the DAQ and trigger stages
  are also shown.}
\label{fig:daq}
\end{figure}

The HERA-B DAQ system and its relationship with the trigger levels is
sketched in figure~\ref{fig:daq}. The trigger rates, latencies and
data volumes of each stage are also shown.

The detector data are read out at the HERA bunch-crossing rate of
about 10 MHz and stored in a 128-deep front-end pipelines during the
First Level Trigger (FLT) processing. The large input event rate
forces the FLT to be entirely built from specialized hardware. The FLT
performs hardware tracking to select events with $J/\Psi$ particles
decaying into two leptons. The FLT tracking is initiated by lepton
candidates in the electromagnetic calorimeter and muon systems. The
accepted events are pushed into a distributed system of buffers
(Second Level Buffers, -SLB-).  The events reside in the SLBs while
the Second Level Trigger (SLT) step is being run.  The SLT is
implemented as a software trigger running in a PC farm of 240 nodes
\cite{dam}.

A switching network provides full connectivity between the SLBs and
the SLT nodes.  This high bandwidth and low latency switch is built by
interconnecting several hundreds of Digital Signal Processors (DSP)
between the SLB system and the SLT nodes.  The total bandwidth of the
DSP switch is above 1 GB/sec. The switch message passing software
ensures zero packet loss and, in addition, possesses multicasting
capabilities which are used for distributing data sets to all SLT
nodes in parallel.

The SLT operates on regions of the detector defined either by FLT
track candidates or pretrigger information (Region of Interest -RoI-).
The SLT refines the FLT tracks and extrapolates them through the
spectrometer magnet, tracks them through the SVD and optionally
performs a vertex cut. Tracker and SVD data needed by the SLT are
fetched from the SLBs via the DSP switch. Events accepted by the SLT
are assembled and, optionally, further processed by the Third Level
Trigger (TLT) in the same trigger node. Events passing the TLT are
sent via a switched Ethernet network to a second 200-processor PC farm
to be fully reconstructed online \cite{gellrich}.  Event
classification by physics category is performed after the event
reconstruction and an additional Fourth Level Trigger (4LT) step can
be run at this stage if further reduction of the event rate is
required.

The HERA-B trigger system relies on track-finding to an unusual
degree.  In turn, accurate track-finding and event selection at all
trigger levels requires relatively precise knowledge of detector
calibration, alignment and detailed detector channel status
information all of which can influence both trigger efficiency and
trigger rate (and therefore system deadtime).
The highly distributed DAQ and trigger systems require a dedicated
online system for monitoring and distributing updates of the CnA
constants into the trigger processors without incurring significant
deadtime.

Running full reconstruction online imposes constraints on the quality
of the CnA constants but also allows immediate data analysis and
therefore detailed information for data quality monitoring. Given the
large data volume collected by the experiment and the large event
reconstruction time, offline data reprocessing should be minimized.

\section{Architecture}
\label{s:design}

\begin{figure}[htp]
\centering
\includegraphics[width=0.9\textwidth]{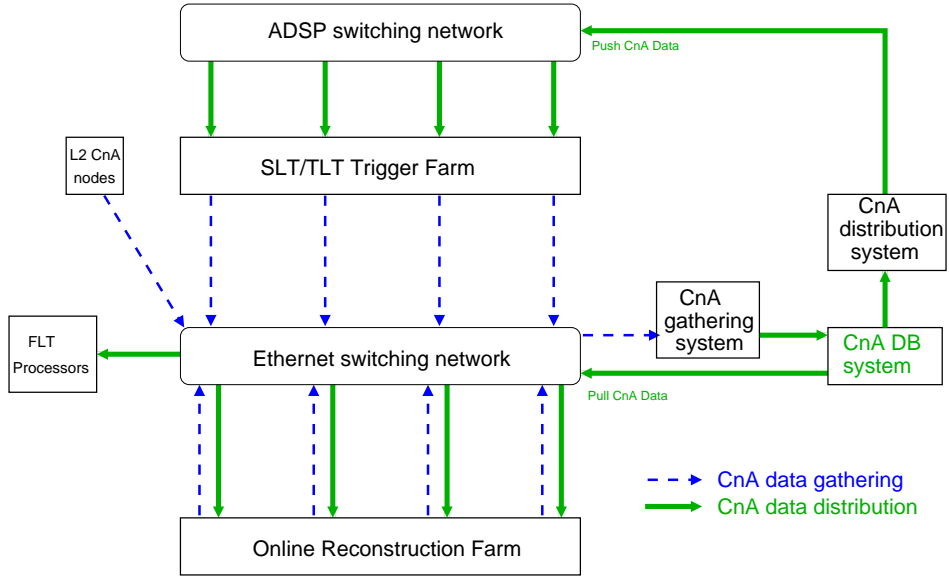}
\caption{Gathering of CnA data and distribution of updated CnA constants.}
\label{fig:cna1}
\end{figure}

The CnA system provides the online infrastructure for collecting data
suitable to align and calibrate the detector, for computing CnA
constants and for delivering updated constants to all processes
involved in the trigger and the online reconstruction, see
figure~\ref{fig:cna1}.  The CnA system takes care of tagging the set
of CnA constants being used at any moment by the DAQ providing an
exact history of the set of constants used.

\subsection{Data gathering}

During the online reconstruction procedure, data for monitoring and
calculating detector conditions are collected from the reconstruction
processes.  In addition, subsystem specific monitors continuously
check the raw data and derive channel status maps.  To make use of the
large number of trigger and reconstruction nodes providing such data
in parallel, the CnA architecture relies on a gathering system to
collect data in a central place. Gathered data can then be used
centrally to compute updated CnA constants which are subsequently
stored in the online database system \cite{amorim}.

\subsection{CnA distribution}

The CnA distribution system delivers updated CnA constants to the
trigger and reconstruction processes.  This involves distributing
large objects to a large number of clients as quickly as possible to
minimize deadtime.  Two different approaches were followed according
to the different latency of the trigger levels and the bandwidth of
the DAQ at those stages (see figure~\ref{fig:cna1}): a push
architecture is best suited for distributing the CnA constants to the
FLT and SLT/TLT processes while a pull architecture was chosen for the
reconstruction farm.

For the SLT/TLT, trigger latencies are of the order of milliseconds and
a fast distribution is required in order not to cause deadtime. Taking
advantage of the high speed, reliability and multicasting capabilities
of the DSP switch, the CnA data can be synchronously pushed to the
trigger processes.

For the online reconstruction farm, several factors favor a pull
architecture. The Ethernet switching network of the reconstruction
farm has substantially less bandwidth than the DSP switch and would be
rapidly saturated if operated under a synchronous push protocol. This
would lead to frequent data retransmissions and consequently, to
degraded performance. In addition, the Ethernet switches have no
support for multicasting so that the same data sets would have to be
sequentially pushed into all nodes.  Furthermore, the online
reconstruction latency is much larger than that of the SLT/TLT and
pausing data taking to wait until all events are fully consumed would
cause deadtime on the order of seconds. On the other hand, the
reconstruction nodes process events asynchronously and independently
from each other, and therefore an asynchronous pull protocol would
distribute the requests for data over the average reconstruction time
and thus make more efficient use of the available bandwidth.  Finally,
with a pull architecture, a distributed system of fast memory database
caches can be implemented to replicate the CnA data and allow for
faster uploading and reduction of overall bandwidth requirements. 

Since uploading into the reconstruction nodes is asynchronous, the
reconstruction processes need to be individually notified when new
constants become available.  The notification is done through the
event data.  A data base table (the "key table") contains identifiers
to all sets of CnA tables which are used by all online processes.  The
identifier of the current key table (the CnA key) is stamped into
events by the SLT process at event assembly time. This index allows an
event to be associated with all the calibration and alignment data
used in its triggering process and online reconstruction. Whenever
updated CnA constants have been distributed to the SLT/TLT nodes or
become available for the online reconstruction, a new identifier is
stamped in the event data. The reconstruction processes check the CnA
identifier and request updated CnA constants when the identifier
changes.

For the synchronous distribution of the CnA constants to the FLT and
SLT/TLT, a manager process is needed for synchronization during the
distribution and also as an intermediary between the CnA producers
(processes producing online updated constants) and the consumers
(trigger and reconstruction processes).  The manager is notified 
when updated constants are available for distribution.
On notification, the manager requests that the FLT/SLT/TLT be paused,
supervises the distribution, and requests resumption data taking
when the distribution is completed.

The system is quite flexible in that it allows distribution of any kind
of information to the trigger processors. This includes FLT and SLT
trigger settings as well as geometry and detector calibration data
sets.  The same distribution protocols used for the on the fly
distributions of CnA constants are employed for the initial loading of
the constants at DAQ booting time.

\subsection{CnA offline usage}

The trigger and online reconstruction farms together with the online
booting, control, monitoring and online data transmission protocols
and processes are used offline for performing data reprocessing and
Monte Carlo production during DAQ idle time \cite{jhnim}.  The CnA
system was also designed for offline use. During data reprocessing,
the CnA system allows any online changes of the CnA conditions to be
accurately reproduced. Moreover, it provides for use of recalculated
sets of CnA constants in place of the online tables, when appropriate.
For Monte Carlo reconstruction, the geometry, calibrations and channel
maps of the run period being simulated are identified and loaded as
well as additional data sets containing detector resolution and
efficiency data.

\section{Implementation and performance}
\label{s:desc}

We describe in this section the implementation of the online
calibration and alignment system following the requirements and
architecture discussed in the previous sections.  Key elements of the
system are the data collection and monitor processes (gatherers), the
distributed system of database servers and proxies for storage and
replication of the constants, and the procedure for distribution to
the trigger and reconstruction processes. The CnA framework also
includes the software modules for the trigger and reconstruction codes
needed to upload the CnA constants.

\subsection{Data gathering and computation}

Data needed to monitor detector conditions are collected, by
gatherers, from the reconstruction processes and from dedicated nodes
in the SLT farm which run subsystem-specific monitors. The dedicated
SLT nodes receive unbiased events at rates up to several Hz and
continuously check the raw event data, updating channel status maps as
needed.

As sketched in figure~\ref{fig:cna1}, gatherers collect summary data
via Ethernet in parallel from all farm nodes. Gatherer processes can
work in two distinct modes, either requesting data from the providers
or subscribing for the data in the provider nodes which then
periodically publish the data to the subscribers.  Gatherers can also
serve as data providers to other gatherers which subscribe to the
provider gatherer for needed data and use the data to update CnA
constants.  In order to limit the amount of CnA data kept locally in
the provider nodes, the CnA data are stored either as histograms or as
ring buffers with arbitrary format.  A remote histogramming package
(RHP\cite{schwanke}) was developed for the data definition and data
collection.  RHP implements part of the functionality of the CERN
HBOOK package \cite{hbook}.

By calling subdetector specific functions, CnA gatherers compute new
CnA constants and monitor their evolution over time.  If significant
changes are produced, the new constants are stored in the distributed
database system described in section~\ref{s:db}, triggering the online
on-the-fly distribution as described in section~\ref{s:dist}.

\subsection{CnA distributed database system}
\label{s:db}

The storage of updated CnA constants into active database servers
triggers online distribution. Upon an update, the CnA database servers
propagate update messages notifying the update to the CnA distribution
system. Indexed objects (CnA keytables), whose identifiers (CnA key)
are stored in the event data, are created automatically by a dedicated
CnA database server.  The CnA keytable contains CnA metadata, namely
the indices of the sets of CnA constants used online during a
particular period. The CnA key allows to associate every event with
all the CnA constants used in its triggering process and online
reconstruction. This bookkeeping is of crucial importance for
identifying the correct sets of CnA constants in the event simulation
and in the offline event data reprocessing as explained in
section~\ref{s:mc}.

The CnA distributed system of databases consists of active subdetector
CnA database servers, a dedicated CnA keytable database server and a
distributed system of fast memory database cache servers.  The
subdetector CnA database servers store the subdetector specific CnA
constants and notify the CnA keytable server of any update. The CnA
keytable server holds the keytable CnA metadata.  After an update
message, it generates a new keytable incrementally, i.e., it copies
the last keytable and updates the indices of the updated sets of CnA
constants. It then publishes to the CnA distribution system the CnA
key of the new CnA keytable.  The distributed system of memory
database caches is used to replicate the CnA constants in order to
speed up their distribution to the online reconstruction farm as
explained in the next section.

\subsection{CnA distribution}
\label{s:dist}

\begin{figure}[htp]
\centering
\includegraphics[width=\textwidth]{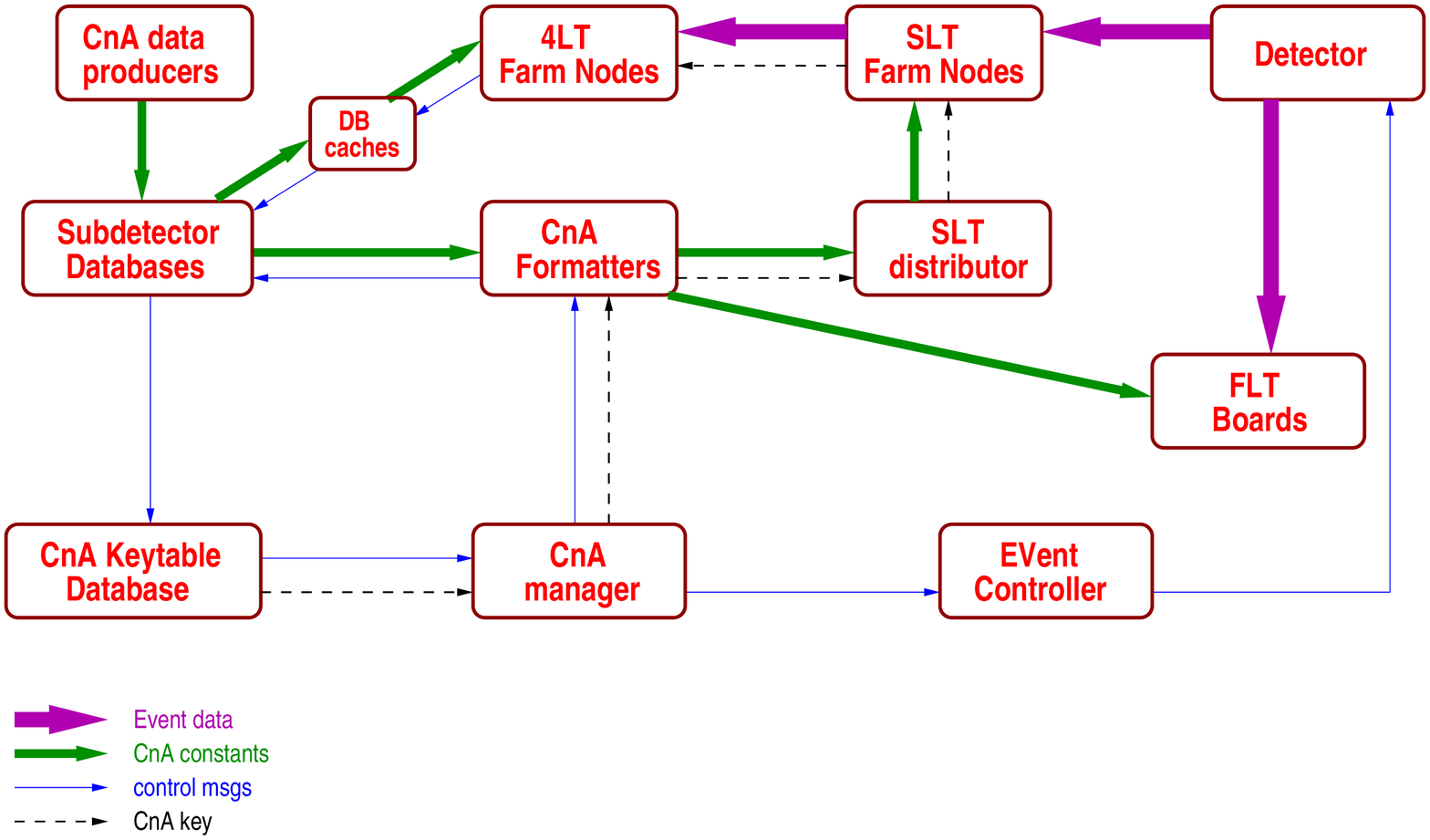}
\caption{Online distribution of updated CnA constants.}
\label{fig:cna2}
\end{figure}

As stated before, the distribution procedure of CnA constants is
initiated by the storage of new constants into active database servers
which then propagate the updates to the CnA keytable database server.
This server in turn notifies the CnA distribution system of the
existence of updated CnA constants.  The distribution procedure is
sketched in figure~\ref{fig:cna2}.

The CnA manager is the process receiving the notifications from the
CnA keytable database server. This process is in charge of the control
and synchronization of the distribution of the CnA constants. At DAQ
booting time, the CnA consumers (trigger and reconstruction processes)
subscribe in the CnA manager for updates of particular sets of CnA
constants.

The format of the CnA constants as stored in the databases might be
different to the format required by the CnA consumers.  To centralize
the formatting of the constants, to save CPU processing time, and to
simplify code in the CnA consumers, CnA formatters were introduced.
The formatters are CnA processes that call subdetector specific
functions which use one or more raw data base tables to produce the
tables which are actually distributed to the consumers.  The
subscription messages sent by the CnA consumers to the CnA manager
identify the associated CnA formatter of the desired set of constants.
The CnA formatters fetch any needed CnA tables from the database, make
the required formatting and send the formatted constants to the CnA
consumers.

The full sequence of the distribution of the CnA constants to the
trigger processors is the following: a CnA producer (CnA gatherer)
stores updated constants into an active database.  The database server
informs the CnA keytable server of the update which creates a new
keytable with a new CnA key index that is sent to the CnA manager. The
manager informs appropriate formatters of the update which then fetch
the updated CnA constants from the appropriate database and produce
format tables. When the formatting is finished the CnA manager
requests the DAQ event controller (EVC) to pause the data taking.  The
EVC waits until all events in the second level buffers are processed
by the SLT nodes before handshaking Cna manager's request. This way
the events already processed by the FLT are processed by the SLT and
are stamped with the correct CnA key. In addition, consuming all
events in the buffers guarantees that the DSP switch bandwidth will be
fully available for the transmission of the CnA constants.  The CnA
manager then requests the CnA formatter to push the constants into the
trigger nodes. The new CnA key is also distributed to the nodes and
will be written into the event data of the new events. When the
distribution is complete, the CnA manager requests the event
controller to resume the data taking.

For distribution to the FLT processors, the FLT CnA formatter sends
the constants through Ethernet to a master process running in each of
the FLT trigger crates which in turn distributes them in parallel to
the trigger boards in the crate. For the distribution into the SLT/TLT
farm, the SLT/TLT formatter sends the constants via Ethernet to a
dedicated process running in one of the SLT/TLT nodes, the SLT
distributor. This process in turn uses the multicasting capabilities
of the DSP switch to transmit the constants in parallel to all the 240
trigger nodes. The throughput of the multicasting via the DSP switch
is about 1 GB/sec.  The constants are synchronously pushed into the
trigger nodes under the coordination of the CnA manager.  Thanks to
the large bandwidth of the DSP switch, the distribution introduces no
significant deadtime.

Unlike the distribution to the trigger nodes, where a synchronous
distribution is done using a push architecture, the distribution of
updated CnA constants to the reconstruction nodes is done
asynchronously in each node and using a pull architecture. After any
update of the CnA constants, the new CnA key index will end up in the
event data. The reconstruction processes upon a change of this index
will fetch from the CnA keytable database the new keytable to
determine which updated set of constants should be loaded. In order to
speed up the loading of the new constants, the reconstruction nodes
fetch them via a distributed system of memory database caches.  Given
the smaller Ethernet bandwidth compared to the DSP switch bandwidth,
an asynchronous retrieval of the constants is more efficient.

Exactly the same distribution procedure is applied at DAQ booting time
to upload the CnA constants into the trigger and reconstruction
processors. At booting time, all the sets of constants appear to be
updated and all of them are distributed. Not only detector conditions
constants are uploaded following the procedure described above, but
also the trigger settings are distributed into the trigger nodes in
the same way.

\subsection{Performance}

Over 60 sets of CnA constants, amounting a total volume of 6.5 MB, are
used. At DAQ booting time, they are pushed into the SLT/TLT trigger
nodes in 1.5 secs, at a rate of 1 GB/sec (6.5 MB x 240 nodes / 1.5
sec). Upon receiving their first events after run startup,
reconstruction nodes fetch the constants at an effective rate of 50
MB/sec (6.5 MB x 200 processes / 25 sec).  Note that in this case, 25
seconds is the total time required to completely upload the constants
in all the reconstruction processes, but not the deadtime caused since
the retrieval is asynchronous and independent in every node so that
each node starts processing events as soon as all the constants have
been read.

The deadtime caused in the data taking by the distribution of CnA
constants to the SLT/TLT nodes is dominated by the time needed for
data transmission and multicasting in the DSP switch. The distribution
messages containing individual sets of CnA constants are multicasted
within the DSP switch. The multicast is based on message copy. The
copy of the messages in the first block of the switch dominates the
multicast latency.  The contribution from the CnA control and
distribution protocol is small, of the order of 100 msecs.

The RICH detector calibration and channel status map were updated
online at intervals of the order of minutes. ECAL pedestals and
tracker channel maps online updates occured at a intervals of the
order of hours.  Although the online CnA system was fully functional,
not all subdector groups developed online monitors or updated CnA
constants online due largely to lack of manpower.  Calibration and
alignment constants for the vertex detector, tracker and muon systems
were updated offline by manually updating the index to the relevent
tables in the online keytable.  The change in the online keytable
caused the new constants to be automatically loaded at the next DAQ
startup or triggered their distribution if data taking was in
progress. 

The FLT lookup tables are distributed to the FLT trigger boards at DAQ
booting time using the CnA distribution procedure. The large number of
these tables and the slow input link to the boards prevents online
distributions of updated lookup tables except as part of the startup
procedure. Steering parameters for the FLT processes are also
distributed via the CnA system.

\section{CnA system for offline reprocessing and Monte Carlo reconstruction}
\label{s:mc}

As the knowledge of the detectors improves, the reconstruction
packages are further developed and improved calibration and alignment
constants are made available, the offline reprocessing of the event
data becomes necessary.  At HERA-B the trigger and online
reconstruction farms together with the online booting, control,
monitoring and online data transmission protocols and processes are
used offline for performing data reprocessing and Monte Carlo
reconstruction as described in \cite{jhnim}.  
Exactly the same CnA distributed database system and CnA data
uploading mechanism used for online reconstruction are also
employed offline for mass data processing. The only difference
between reprocessing and online reconstruction is that the source of
the data is not the detector but the recorded raw events archived on
tape. This system makes an extremely efficient use of the online
computing resources during idle time and shutdown periods of the
detector.

As mentioned earlier, the event record includes a tag which
links the event with the CnA keytable in the database containing 
the indices of all the sets of CnA constants used in the online 
reconstruction of that event. 
The automatic bookkeeping of the keytables in the database 
during data taking allows to reproduce the detector calibration and
alignment conditions for offline data reprocessing. Sets of constants
improved offline are incorporated in the reprocessing by producing
a revision of the online keytables. The online keytables are first
duplicated in the database with a new revision number and then the
keytables corresponding to data taking periods for which updated
constants are available are modified with the indices of the updated constants.
The offline reconstruction
processes make use of a given revision number of the keytables
when reprocessing the data. 

Monte Carlo event reconstruction should be performed using the 
reconstruction conditions of the real data one
intends to simulate. This is simply achieved by using the keytable (CnA
key and CnA revision number) employed in the reconstruction of the real
data. For extended data taking periods with several associated keytables,
Monte Carlo samples can be reconstructed using those keytables separately, and the 
events are reweighted according to the relative luminosities of the periods
of validity of the different keytables.

\section{Summary}

In the HERA-B experiment, all trigger levels as well as the online
reconstruction critically depend on calibration and alignment
constants. In order to keep the trigger performance and the online
reconstruction stable under variations of the detector conditions, an
online calibration and alignment system was implemented and used. This
system monitors the status of the calibration and alignment constants,
recomputes them upon significant changes in the calibration or
alignment conditions in the detector and if necessary distributes them
on the fly to the trigger and reconstruction processors without
causing significant deadtime in the data acquisition.  The
distribution system exploits the high bandwidth and multicasting
capabilities of the DSP switch to synchronously push the constants to
the SLT/TLT trigger processes with a throughput of 1 GB/s. On the
other hand, given the smaller effective network bandwidth of the
reconstruction farm and the higher event processing time, the
reconstruction processes asynchronously fetch the updated constants
from a distributed and replicated database system.

A tag in the event record associates every event with the detector
conditions used in the trigger and online reconstruction. This
mechanism provides the bookkeeping necessary for offline data
reprocessing and Monte Carlo reconstruction. The online CnA
distribution system is also used offline for mass data processing.

The integration of the CnA system took place during the HERA-B
commissioning runs in 2000/2001. The system was fully operational and
routinely working during the 2002-2003 data taking period and is still
in use for data reprocessing and Monte Carlo reconstruction.

The upcoming LHC experiments will incorporate PC farms into their DAQ
and trigger systems and might find the HERA-B experience concerning
the online calibration and alignment system of interest.

\section{Acknowledgments}
We are grateful to Andreas Gellrich for fruitful discussions. We thank
the DAQ subdetector and trigger experts for their work in the
integration of the online subsystems into the DAQ CnA framework.

\bibliographystyle{elsart-num}
\bibliography{cna}

\end{document}